\documentstyle[aps,preprint,tighten,floats,epsf,rotate]{revtex}

\begin{document}
\draft

%
\title{Probing Gauge String Formation in a Superconducting 
Phase Transition}
\author{Serge Rudaz \footnote{e-mail: rudaz@mnhep1.hep.umn.edu}}
\address{School of Physics and Astronomy, University of Minnesota,
Minneapolis, Minnesota 55455, USA}
\author{Ajit M. Srivastava \footnote{ajit@iopb.res.in} and
Shikha Varma \footnote{shikha@iopb.res.in}}
\address{Institute of Physics, Bhubaneswar 751005, India}
%
%
\maketitle
\widetext
\parshape=1 0.75in 5.5in
\begin{abstract}
 Superconductors are the only experimentally accessible systems
with spontaneously broken gauge symmetries which support topologically
nontrivial defects, namely string defects. We propose two experiments 
whose aim is the observation of the dense network of these strings
thought to arise, via the Kibble mechanism, in the course of a spontaneous 
symmetry breaking phase transition. We suggest ways to estimate 
the order of magnitude of the density of flux tubes produced  in the 
phase transition. This may provide an experimental check for the 
theories of the production of topological defects in a spontaneously
broken gauge theory, such as those employed in the context of the
early Universe.
\end{abstract}
\vskip 0.125 in
\parshape=1 -.75in 5.5in
\narrowtext

\centerline {\bf 1. INTRODUCTION}

 Formation of topological defects in phase transitions has been of great
interest to condensed matter physicists, as well as to particle 
physicists in the context of the early Universe \cite{vs1,vs2}. 
Production of topological defects in phase transitions is conventionally 
estimated using either thermal production (with the usual Boltzmann 
suppression of defect density) \cite{thrm}, or via the {\it Kibble 
mechanism} which dominates over the thermal production in a variety of 
transitions, and arises due to the formation of a domain structure 
\cite{kbl}. Recently, a new mechanism has been proposed for production 
of defects where defect-antidefect 
pairs are produced due to energetic oscillations of the magnitude of the 
order parameter field \cite{dgl}. Though superconducting transition is 
dominated by dissipation (which suppresses field oscillations), the 
presence of gauge fields may make this new mechanism effective in this 
case also, see ref. \onlinecite{cs}. For the purpose of this paper,
we will use estimates of string production based on the Kibble mechanism,
though actual string distribution may have contributions from this new
mechanism as well.  

 In the Kibble mechanism, the order parameter is taken to be roughly 
uniform within a correlation region (domain), while varying randomly
from one domain to the other. In between any two adjacent domains, 
the order parameter field is supposed to vary with least gradient 
(this is usually called the $geodesic~rule$). Consider a superconducting
phase transition corresponding to spontaneous breaking of U(1) gauge
symmetry and characterized by a complex order parameter. Magnitude
of the order parameter gives the degree of superconductivity, while 
its phase $\theta$ can vary spatially over distances larger than
the coherence length. In this case, string defects arise
at the junctions of domains if $\theta$ winds non trivially around a 
loop going through adjacent domains.  Simple arguments show 
\cite{smln,nlc} that the probability of string formation through a 
triangular domain (in two space dimensions) is equal to 1/4. 

 It was first suggested by Zurek \cite{zurk1} that one may be able to 
experimentally test the Kibble mechanism in superfluid $^4$He system. 
Later, an experimental verification of the Kibble mechanism was carried 
out in a sample of liquid crystals where strings arise due to formation
and coalescence of nematic bubbles in the isotropic phase \cite{nlc}, see 
also ref. \onlinecite{turok}. Subsequently, experimental tests of Kibble 
mechanism have been carried out in superfluid helium systems \cite{he43}. 
Recently, prediction of correlations in defect-antidefect production via 
Kibble mechanism has also been experimentally verified in liquid crystals, 
see ref. \onlinecite{crln}. 

  An important limitation of these experimental studies 
\cite{nlc,turok,he43,crln} is that they all correspond to the formation of 
global defects. Some of the most important types of topological defects
in the context of particle theory models of the early Universe are
local defects, i.e. those which arise due to spontaneous breaking of 
a gauged symmetry. Examples of local defects are magnetic monopoles,
and many types of cosmic strings. For local defects, the application of 
Kibble mechanism becomes somewhat  non-trivial. For example, it was 
argued by two of us \cite{gdsk} that the geodesic rule becomes ill 
defined in the case of gauged topological defects.
For global defects, the geodesic rule is well
motivated as from energy consideration we expect least gradient of order 
parameter between two domains. However for gauged defects such energy
considerations are absent, as spatial gradient of the order parameter 
alone, in between two different points in space, does not have any 
physical meaning. It was shown later in refs. \onlinecite{brnd,kv} 
that for certain situations, the geodesic rule arises dynamically in 
gauge theories. Recently, Copeland and Saffin have studied \cite{cs} 
collisions of bubbles in a first order transition case for
Abelian Higgs model, and have shown that geodesic 
rule can get violated due to field oscillations, (see also 
ref. \onlinecite{dgl} in this context).

It does seem, therefore, that the production of local defects may show
qualitatively new features as compared to the case of global defects. Thus 
it becomes important to investigate whether experimental studies such as 
in refs. \onlinecite{nlc,turok,he43,crln} can be extended to systems
with local defects. We consider this issue in this paper. The only system
known to actually occur in nature with local defects is that of flux tubes
in type II superconductors. [We note that it is known that certain types 
of liquid crystals resemble gauge systems \cite{gennes}. It will be very 
interesting to study the defect production in this class of liquid 
crystals.] It has been suggested by Zurek \cite{zurk2} that superconducting 
transition in a torus geometry may provide a good way to detect domain 
formation. However, one can not probe the string distribution directly 
in this manner. In Sect. 2, we describe our proposal for one experiment 
which is aimed to directly  detect open string segments exiting the surface
of a superconductor. This provides a possible test of defect formation 
in two space dimensions. In section 3, we describe another experiment to 
detect the formation of the full 3-dimensional network of the strings 
formed inside the sample. 
 
\vspace*{8mm}
\centerline {\bf 2. DETECTION OF OPEN STRING SEGMENTS}
\vspace*{4mm}

 Consider a superconducting phase transition producing a string network 
as shown in Fig.1a. The arrows on the strings denote the direction of 
the magnetic field.  Top surface of the sample is shown in Fig.1b with
a $+$ sign denoting a string exiting the surface while a $-$ sign 
denoting a string entering the surface. If strings and antistrings 
form independently then there will be statistical fluctuations leading 
to an excess, of order $\sqrt N$, of one kind of string over 
the other, where $N \sim A/(4\xi^2)$ is the total
number of all strings ($A$ being the surface area).
However, as pointed out in ref. \onlinecite{smln}, 
the net string number through a surface bounded by a loop of perimeter 
{\it L} is proportional to $\sqrt {\it L}$. Basically the argument is that 
the length ${\it L}$ of the perimeter consists of $n_{\it L} 
(= L/ \xi)$ segments of size correlation length ($\xi$) each  and 
$\theta$  varies  randomly beyond each segment. This, then, is
a random walk problem with average step size equal to $\pi/2$
(since the smallest step size for $\theta$ is equal to zero while 
the largest step size has magnitude $\pi$). The typical value of
the net winding (i.e. net increase in $\theta$ divided by $2\pi$) along 
the perimeter will then fluctuate about zero with typical width given 
by  $n_{ex} \simeq \sqrt n_{\it L}/4$. This suppression in net winding 
holds for other topological defects as well \cite{ams} and arises since 
the presence of a string  in a given domain increases the probability of 
the formation of an antistring in the adjacent domain due to partial 
winding of the phase $\theta$ along the boundary common to the 
two domains. This correlation between the formations of defects and
antidefects has been recently experimentally verified for global
defects in nematic liquid crystals, see ref.\onlinecite{crln}. 
 
This excess string number $n_{ex}$ will then lead to a net magnetic flux 
through the surface with magnitude 

\begin{equation}
\phi_{net} \simeq {\phi_0 \over 4} \sqrt{{L \over \xi}} ~ .
\end{equation} 

  Here $\phi_0 = {hc \over 2e}$ (= $2 \times 10^{-7}$ Gauss.cm$^2$) is 
the flux quantum. Note that it may  not be appropriate to associate 
a single flux quantum with each vortex since these vortices will be 
strongly overlapping.  However, the excess vortices (or anti-vortices) 
will be very dilute and $\phi_{net}$ should be obtainable by assuming 
that these excess vortices each carry a single flux quantum. 

 For the domain size, one usually takes the correlation length at the 
Ginzburg temperature $T_G$, since for temperatures above $T_G$ 
thermal fluctuations can make defects unstable \cite{kbl}. However, large 
scale structure of string distribution may remain unaffected by such 
fluctuations, as recently argued by Kibble (see, ref. \onlinecite{vs1}), 
and by Zurek \cite{zurk2}. They have further argued that the appropriate 
value of $\xi$ should depend on the rate of phase transition \cite{vs1}.  
For example, it has been argued in ref. \onlinecite{zurk2} that, due to 
critical slowing down, the order parameter may be frozen above $T_c$ 
itself.  For a superconducting phase transition, the corresponding  
frozen out correlation length is given by \cite{zurk2},

\begin{equation}
\xi_f = 10^{-2} ({\xi_0 \over 1000 \AA}) \tau_Q^{1/4} ~cm.
\end{equation}

 Here, $\xi_0$ is the zero temperature correlation length and $\tau_Q$,
measured in seconds, is the time scale of the quench (assuming that the 
relative temperature $(T_c-T)/T_c ~ \simeq t/\tau_Q$, $t$ being the time).
Clearly, $\xi_f$ can be much larger than the value of $\xi$ at 
$T_G$. In the present paper, we will use $\xi_f$ to be the 
appropriate correlation length. Though we mention that the issue of the
appropriate correlation length for defect formation is not a settled one.
Thus it is important to emphasize that all our conclusions about the
experimental proposals in this paper will be straight 
forwardly applicable if the correlation 
length happens to be smaller than the value given by Eq.(2). In fact, 
smaller correlation length will make the experiments much easier.

 As $\phi_{net}$ is larger for smaller $\xi_f$, High T$_c$ superconductors 
should be excellent for this purpose as they have very small 
coherence length (with $\xi_0$ of the order of 10 $\AA$). [Though,
the value of $\phi_{net}$ may then also depend on the surface chosen 
due to the anisotropic magnetization in high T$_c$ materials, see 
\cite{anis}.] Assuming a very rapid quench, (presumably by taking very 
thin layer of superconductor and rapidly cooling it
with $\tau_Q$ of order of 1 sec. to about 10 milli seconds) one may get 
$\xi_f \simeq 0.3 - 1$ $\mu$m.  With $\xi_f \simeq 0.3$ $\mu$m, 
Eqn.(1) gives a magnetic flux of the order of 50 - 90 $\phi_0$ for  a 
square sample with 1 cm sides which should be easily measurable using 
SQUID.  Also, the number of strings and antistrings $N$ through the 
sample is  about 10$^{8}$. As flux tubes are generally strongly pinned 
for high $T_c$ superconductors, one may be able to use STM techniques 
to directly probe these flux tubes exiting the surface \cite{elec}.

 As suppression in $\phi_{net}$ arises from
the correlation between string and antistring formation, it 
suggests that if we somehow disrupt this correlation then 
a larger value of $\phi_{net}$ can be achieved. For example, 
if a sample consists of $n_g$ isolated grains, then string-antistring
correlation will be absent in-between two different grains. The typical
net flux through the whole sample will now be $\simeq \phi_{net}
\sqrt{n_g}$ where $\phi_{net}$ is typical flux through a single grain. 
It is important to note that for such a sample one can not just 
calculate the net increase in the phase $\theta$ along the perimeter 
of the whole sample as we did before, since that required the use of 
the {\it geodesic rule} to prescribe $\theta$ between two domains. 
However, now for two isolated grains, $\theta$ is not defined 
in-between the grains.

 For a grain of about  3 $\mu$m size (as in ref. \onlinecite{trp}), the 
net flux through its surface will be about 1 - 2 $\phi_{0}$, assuming 
$\xi_f \simeq .3 - 1$ $\mu$m (with the total number of strings through 
the grain's surface being about 2 - 25). A square sample with sides 
equal to 1 cm will 
contain about $10^7$ grains, leading to $\phi_{net} \simeq 3 - 6 \times 
10^3 \phi_0$.  We should mention here that for very small grains flux 
creeping may be a serious problem. Also, if the size of each grain is 
comparable to (or smaller than) $\xi_f$ then 
one will not expect even a single
flux tube to be formed. In fact, this can provide a nice way to determine
the value of $\xi_f$ appropriate for superconducting transition (occurring
in a given time scale). By taking samples, consisting of grains 
of different sizes, one can find the size when flux tube formation
abruptly falls to zero. The corresponding size of the grains will be
of the order of $\xi_f$.

 Note that $\phi_{net}$ as given in Eqn.(1) measures just the difference 
between strings and antistrings, and hence will not change due to 
string-antistring annihilations. $\phi_{net}$ can change due to 
migration of open segments out of the boundary of the sample and should
lead to a fractional correction of the order of 
$1/\sqrt A$ (for short times after the transition).
Here we mention an important point that during the early stages
of the formation, vortices having only winding number of the order
parameter without any associated magnetic field, may have a faster creeping
rate compared to the creeping of flux vortices. The 
effects of vortex creep can be reduced by taking large sample area  
to increase the vortex creep time while reducing the phase transition
time by taking very thin sample. $\phi_{net}$ can also change due to 
strings which enter from the top surface but exit from the side surface
(instead of the bottom surface). These strings may then shrink quickly and 
disappear through the edges of the sample. This, however, will lead to a 
fractional correction in $\phi_{net}$ of the order of the ratio $r_s$ of 
the area of the side surfaces to the area of the top and bottom surfaces,
as strings entering from top will have a probability of $r_s$ to get out 
through the side surface.

 An important thing to realize here is that  for high
T$_c$ materials vortices are generally strongly pinned. This clearly 
suppresses both the above effects and should make high T$_c$ materials
best suited for this experiment.

 We mention here the work in ref. \onlinecite{trp} where flux trapping was
observed in high T$_c$ materials with varying external magnetic
field. It was observed in ref. \onlinecite{trp} that for extremely small
applied field the magnetization in the sample was consistent with
zero, perhaps suggesting suppression of string formation. 
The high T$_c$ material used in ref. \onlinecite{trp} was in the form
of a fine powder with grain size of about 3 $\mu$m. First, 
for such small samples, flux creeping (or, more importantly, vortex
creeping) may be crucial. This is especially so when we note that for 
such grains the area of all surfaces may be of similar size which will 
then lead to shrinking of many strings through the edges. [This 
shrinking of strings can be avoided by taking grains of very small 
thickness.] It is also possible that the phase transition in that work 
may not have been carried out sufficiently fast, so the value of 
$\xi_f$ (Eq.(2)) may have been larger than the grain size. In such 
a situation one will not expect any flux tubes to form.  

\vspace*{8mm}
\centerline {\bf 3. DETECTING STRING LOOPS INSIDE}
\centerline {\bf  THE SAMPLE}
\vspace*{4mm}
 
 Vortex formation in two space dimensions is quite different from that
in three space dimensions. Even though the underlying mechanism
may remain the same, namely the Kibble mechanism, the dynamics of
string network in three dimensions has qualitatively new features.
In two dimensions, vortex-antivortex annihilate leading to decrease
in the number density of vortices in time. In three dimensions, 
string density decreases due to shrinking of string loops, accompanied 
by another important feature, usually termed as the intercommuting 
property of strings. Intercommutation means that when two strings 
cross, they exchange partners at the crossing point (see Fig.2). We 
now describe a proposal for a second  experiment to detect the 
formation of full 3 dimensional string network inside 
the sample, utilizing this intercommutation property of strings. 

  Numerical simulations show \cite{smln}, 
that the average string length passing through a (cubic) 
domain is about 0.88 $\xi$ (for U(1) strings). Thus
the density of strings (string length per unit volume) 
expected is of the order of $\xi^{-2}$. Evolution of the 
resulting  string network happens by shrinking of loops, and
by intercommutation of strings. There are various numerical 
simulations which show that strings (global as well as local) 
intercommute under most generic conditions \cite{shlrd}, see also
ref. \onlinecite{carl} for a qualitative discussion. Intercommutativity 
for global strings has been seen to occur in liquid crystal experiments 
\cite{nlc,turok}. [A proposal for experimentally checking 
intercommutativity of gauge strings by observing the crossing of flux 
tubes in type II superconductors has been discussed in 
ref. \onlinecite{crs}.] Based on the overwhelming evidence for 
intercommutativity of strings, we assume that flux tubes in type II 
superconductors also intercommute under most generic conditions. 
[Recently, there has been some discussion of cutting and reconnection of 
vortex lines in the context of high $T_c$ superconductors \cite{crshtc}.
Though, as we will explain below, due to presence of impurities
high $T_c$ superconductors are not very suitable for this second
experiment.]

  Consider now the experimental setup shown in Fig.3a. A superconducting
slab is surrounded by four electromagnets which are suspended so that
their deflections (in the plane of all the four magnets) can be measured. 
Dashed lines show the magnetic field lines. As the temperature is lowered 
through the transition temperature, the magnetic flux will get expelled 
out, except in regions where it will form flux tubes going through the 
sample (assuming that the magnetic field is (slightly) larger 
than the lower critical field $H_{c_1}$). Along with this there will also
be some strings produced due to the Kibble mechanism (most or all of which 
are assumed to be in the form of closed loops for simplicity). Let us first 
illustrate the basic idea of the experiment by 
considering a simple, idealized case when
only two flux tubes are passing through the sample due to external magnets
and a single large string loop is formed due to Kibble mechanism which 
surrounds the two straight flux tubes, see Fig.3b. If these straight flux 
tubes were not present, then this string loop would have collapsed all the 
way. However, when the straight strings are present then the collapsing
string loop will have to cross these straight strings. [Note that, though
parallel strings can get entangled due to repulsion, a collapsing loop
can not remain entangled.] As soon as this crossing happens, the strings 
are going to intercommute (by tilting of the loop, if required)
resulting in the flux tubes as shown in Fig.3c. These strings will then shrink
down towards the surface of the sample. We see that the final flux tube
distribution in Fig.3d is very different from the initial one as shown
in Fig.3b. The force between the external magnets and the superconducting
sample depends upon the distribution of trapped flux tubes \cite{force}.
As the flux tube distribution has drastically changed from Fig.3b to Fig.3d,
this should lead to different forces on magnets which could be measured.
 
  We mention that the reverse of the process, leading to
a distribution of flux tubes as in Fig.3d evolving back into the one shown
in Fig.3b, is very difficult because the strings shrink due to dissipation 
and for strings to change from Fig.3d to Fig.3b it is necessary that a large 
loop of string forms somewhere which then expands and touches the outer 
surfaces of the sample. Of course the original, thermodynamically stable,
string configuration may get restored due to strings creeping from outside.
However, for intermediate times one will be stuck in the situation where 
the flux tubes have been broken and diverted in different directions. 

 We now discuss, in detail, the realistic case when a large number of 
strings are formed.  Let us first consider the situation when there is no 
external magnetic field and the sample is cooled to a superconducting state.
After waiting for long enough time so that any strings produced during
transition have all disappeared (apart from the possibility of pinned 
vortices), we apply the external magnetic field which will then
lead to a group of straight strings (as shown in Fig.4a). All magnets
are chosen to be of equal strength and the distance between a pair of magnets 
on each side of the sample is chosen to be large so that the two bundles of 
strings are reasonably confined and far away from each other.
Next, one heats the sample back to the normal phase and then again lowers 
the  temperature to superconducting state, now in the presence of the 
external magnetic field. Along with straight strings due to the external 
magnetic field,  many strings will form  due to the Kibble mechanism as 
well, see Fig.4b. We assume for simplicity that all Kibble
strings are closed loops.  As these loops  shrink, many of these straight 
strings  (due to the external magnets) will get broken and diverted  
(similar to the situation in Fig.3c) resulting in a different 
distribution of strings, see Figs.4b and 4c. 

  Let us now consider the question of the observation of these diverted 
strings.  As we mentioned, recent techniques \cite{elec} seem very promising 
in real time observations of strings even though the density and the 
randomness of the string network may pose serious problems. 
Even if the real time observation techniques in ref.\onlinecite{elec} 
can not be used for an initially  dense, fast evolving, 
random network of strings, one may be able to use it to observe the 
final resulting networks of strings at stages shown in Fig.4a and Fig.4c.
Comparison of the two will directly give 
the fraction of strings which have been diverted. 

 Note that the Kibble estimate of strings will be modified for domains
through which flux tubes due to external magnetic field pass.  However,
for a strongly type II superconductor, with a dilute bundle of strings,
the number of domains affected by this will be small. For example, if the 
penetration depth is 5 times bigger than the correlation length then at 
most 4\% domains will be affected in that region.

  There is a simple way to observe diverted strings. This depends on 
the estimate of how many strings are expected to get diverted.
Let us consider the case when the magnetic field $H$ due to the external
magnets is only slightly above the lower critical field $H_{c_1}$,
with $H - H_{c_1} << H_{c_1}$. The resulting flux tube network is
very dilute with the spacing $d$ between the flux tubes (for triangular
flux tube lattice) given by \cite{park},

\begin{equation}
d \simeq \lambda ~ ln[{6 \phi_0 \over 8\pi \lambda^2 (H - H_{c_1})}]
\end{equation}

\noindent where $H_{c_1}$ can be expressed in terms of the coherence 
length $\xi$ and the penetration depth $\lambda$ \cite{park}, 

\begin{equation}
H_{c_1} = {\phi_0 \over 4\pi \lambda^2} ln({\lambda \over \xi}) ~.
\end{equation}

  $\xi$ here should be taken to be the equilibrium correlation length. We 
take the domain size to be again determined by the frozen correlation 
length $\xi_f$ as given in Eq.(2).  [We again emphasize that qualitative 
aspects of our results do not depend on this choice. In fact for smaller 
correlation length, flux tube density will be larger and hence, 
detection of flux tubes will become easier in these experiments.]
The length of strings per unit volume due to Kibble mechanism, $L_s$, 
is expected to be, $L_s \simeq 0.88 \times \xi_f^{-2}$ (see ref. 
\onlinecite{smln}), while
the string length per unit volume due to external field will be
$L_{ext} \simeq d^{-2}$. By varying $H$ from a value very close to $H_{c_1}$
up to a much larger value, we can explore situations ranging from 
$L_{ext} << L_s$ to $L_{ext} >> L_s$. [For the latter situation,
one can use slower phase transition rate to increase $\xi_f$, leading
to lower $L_s$.]

 Now, to estimate the fraction of diverted strings, one approach can be 
to note that at the time of string formation, the number $N_d$ of loops 
with sizes greater than $d$ (for $d$ comparable with the sample size) is 
of order one (from the results 
of numerical simulation in ref. \onlinecite{smln}). This will suggest 
that the fraction of diverted strings will be of order $1/N_{ext}$ 
where $N_{ext}$ is the number of external strings due to one pair of 
magnet. Clearly this fraction will be very small. [Still these 
diverted strings may be observable by Lorentz microscopy \cite{elec},
as we mentioned earlier.] However, we now argue that the fraction of
diverted strings can be much large, even if $N_d$ is very small.
The point is that string loops at the time of formation do not
remain unchanged during the evolution and coarsening of string network
which happens by shrinking and intercommutation of strings. 
First consider the case when external magnetic field is chosen to be
sufficiently close to $H_{c_1}$, so that $L_{ext} << L_s$.
Let us concentrate on the strings due to the external
field. As the strings enter  the sample from outside, they will
very quickly start intercommuting with shrinking  strings in the
highly dense string network and, within few random walk 
steps, the memory of the original direction of entrance (or exit) will be 
lost. This suggests that as the string network thins out due to the 
shrinking of strings, a given string will either follow its original 
straight path or will be diverted to the other magnet, both the
possibilities being equally probable (as all the magnets are chosen to 
have the same strength).  Thus in the limit $L_{ext} << L_s$, 
about 50 \% of the straight strings will be diverted as shown in Fig.4c.

  This is a macroscopic effect. Straight strings going through the sample
result in a force on the magnets \cite{force}.
If 50 \% of those strings get diverted, the force due to 
strings on the magnets will change significantly affecting 
deflections of the magnets (compared to their positions in Fig.4a). 
Observation of such a deflection will signal the presence of strings 
produced in the phase transition with density far in excess of 
$L_{ext}$. We can now increase the external magnetic field $H$ which leads 
to an increase in $L_{ext}$. When $H$ has been increased sufficiently
above $H_{c_1}$ so that $L_{ext} >> L_s$ then a 
very small fraction of straight strings should
be diverted, leading to negligible deflection of the magnets.
The deflection of magnets will therefore increase with increasing value
of ${L_s \over L_{ext}}$ and will saturate at some maximum value
when $L_s > L_{ext}$, corresponding to the situation when about 50 \% 
of strings get diverted. The value of $L_{ext}$ for which the deflection
almost saturates will give a direct measure of the order of magnitude
of the density of strings  produced in the Kibble mechanism.
We note here that it is not important to know what the actual change 
in the force on magnets is when the strings get diverted: The important
thing is that the change in force will increase with decreasing
value of external magnetic field  and will approach saturation when 
$L_{ext}$ and $L_s$ are roughly of the same order of magnitude.
We should point out here that the presence of 
open string segments may change the number of strings which ultimately get 
diverted. It is not difficult to estimate the number of strings which 
are diverted depending on the density of open strings, though its effect
may not be too significant. This is because, $\phi_{net}$ through a surface
of area $A$ is proportional to $A^{1/4}$, and it is $\phi_{net}$ which
can affect number of diverted strings (as strings-antistrings ending
on a surface will eventually join due to attractive force, and then will 
get pulled inside the sample).

 It is important to realize that this experiment requires a very 
clean sample. Thus high $T_c$ superconductors are not suitable for 
this experiment. (Also, for high $T_c$ materials, mean field results
like in Eqn.(3) may be suspect \cite{mnfld} near $H_{c_1}$. Though, it will 
only affect the value of $H$ at which saturation in deflections of magnets 
may happen, without changing any results in this section.) Vortex pinning 
can affect the results of this second experiment in two important ways. 
First, with pinning centers present, the 
magnetization of the superconductor shows 
irreversible behavior. Thus the number of flux tubes present in the sample 
for  external fields applied after the superconducting transition will
be different than the number for the case where the transition 
is carried out in the presence of the external field. For this experiment 
we have assumed the two numbers to be the same. The second problem
will be that with pinning, we can not assume that the strings which
have intercommuted will shrink away to the surface. Even after
intercommuting, the strings may remain pinned inside the sample.
It seems very difficult to make any estimate of how much
impurity can be tolerated  for the success of this experiment. 
If sample is pure enough that the number of pinned strings is much
smaller than the number of diverted strings then pinning should
not significantly affect the results of the experiment.

\vspace*{8mm}
\centerline {\bf 4. CONCLUSIONS}
\vspace*{4mm}

  We conclude by emphasizing the importance of experimentally checking
theories of formation of gauge defects. As we mentioned earlier, gauge
defects play an extremely important role in particle theory models of
the early Universe. Constraints on various theories of Grand Unification
arising from over abundance of magnetic monopoles are well known. Many
models of structure formation are based on production and evolution of
gauge cosmic strings. Due to non-trivial issues relating to the validity
of the geodesic rule in gauge theories, it becomes important to test
defect formation for gauge defects, just as experimental tests have
been done for the formation of global defects in liquid crystals and 
in superfluid helium. The only system known in nature, with spontaneously
broken gauge symmetry, with a vacuum manifold nontrivial enough to 
lead to topological strings is superconductor. The proposals for
the two experiments in this paper, to detect string formation in 
superconductors, thus provide a way to check some of the important 
conceptual issues relating to theories of gauge defect formation. 

 We point out some potential problems in the second experimental 
set up. After the strings get diverted, more strings may eventually creep in 
from the surrounding areas to make the local string density consistent
with the value of the external field, although, once a string configuration
like in Fig.4c has been achieved (due to intercommutativity), there may
be a large activation energy required for strings to creep in (something
like the energy required to create a large string loop). It is important,
therefore, that the two magnets be far apart so that the two bundles of
strings in Fig.4a are far away from each other. The observations of strings
(either by using  Lorentz microscopy or by the deflection of magnets)
will have to be done before this creeping becomes significant. 
Also as we mentioned earlier, the sample need to be extremely
clean as otherwise pinned vortices can make any interpretation of
the results difficult. From this discussion it is evident that the first
experiment should provide a relatively clean way to check the string
formation and a high T$_c$ material can be used for this purpose.
Vortex pinning not only does not cause any problem for this experiment, 
it actually helps by reducing the errors caused by flux creep.
The second experiment, all though requiring very clean sample, provides a 
way to probe the full 3-dimensional network of gauge strings. Understanding
properties of such an evolving string distribution are of crucial 
importance for cosmic string models of structure formation in the
Universe. 

\acknowledgements

 We thank Mathew Fisher for many useful comments and especially for
pointing out the difference between the vortex migration and flux tube
migration, and Tanmay Vachaspati for pointing out an error in our
earlier calculation of net string number through a surface.
We are very grateful to Jim Langer, Carl Rosenzweig 
and Mike Stone for very useful discussions and for comments about
flux trapping experiments and the issue of reversibility of intercommuted 
strings. We also thank Yutaka Hosotani, Fong Liu, Trevor Samols, Sanatan
Digal, and Supratim Sengupta for useful discussions. Earlier part of 
this work was supported by the U.S. Department of Energy under contract 
number DE-AC02-83ER40105 for S.R. and by the U. S. National Science 
Foundation under Grant No.   PHY89-04035 for A.M.S.


\newpage

\vspace*{8mm}
\centerline {\bf FIGURE CAPTIONS}
\vspace*{4mm}

 Fig.1: (a) Formation of a string network in the phase transition in a 
thin slab of superconductor. [The random network we show in these 
figures is only an illustration and is not exactly what one actually 
gets in a simulation.] (b) Distribution of string ends at the top
surface of the sample. $+$ and $-$ denote strings exiting
and entering the surface respectively.
    
 Fig.2: Intercommutation of strings.
    
 Fig.3: A superconducting slab is shown in (a) 
surrounded by four electromagnets which are suspended so that
their deflections  can be measured. 
Dotted lines show the magnetic field lines. (b) shows the
idealized case when only a single large string loop is formed through 
the Kibble mechanism, which  surrounds only two straight flux tubes. 
(c) shows that strings have intercommuted as the loop collapses and crosses
the straight strings. The final flux tube
distribution is shown in (d) which is  very different from the 
initial one as shown in (b).
    
 Fig.4:  Bundles of strings due to the external field in the 
superconducting state obtained by switching on the external field long 
after the superconducting transition is completed (so that any strings 
produced during the transition have disappeared. (b) shows
the network of strings expected after cooling the sample to
superconducting state in the presence of the magnetic field of
the external magnets.  (c) shows the diverted strings due to string
intercommutation.

\newpage

\begin{figure}[h]
\begin{center}
\vskip -1 in
\leavevmode
\epsfysize=18truecm \vbox{\epsfbox{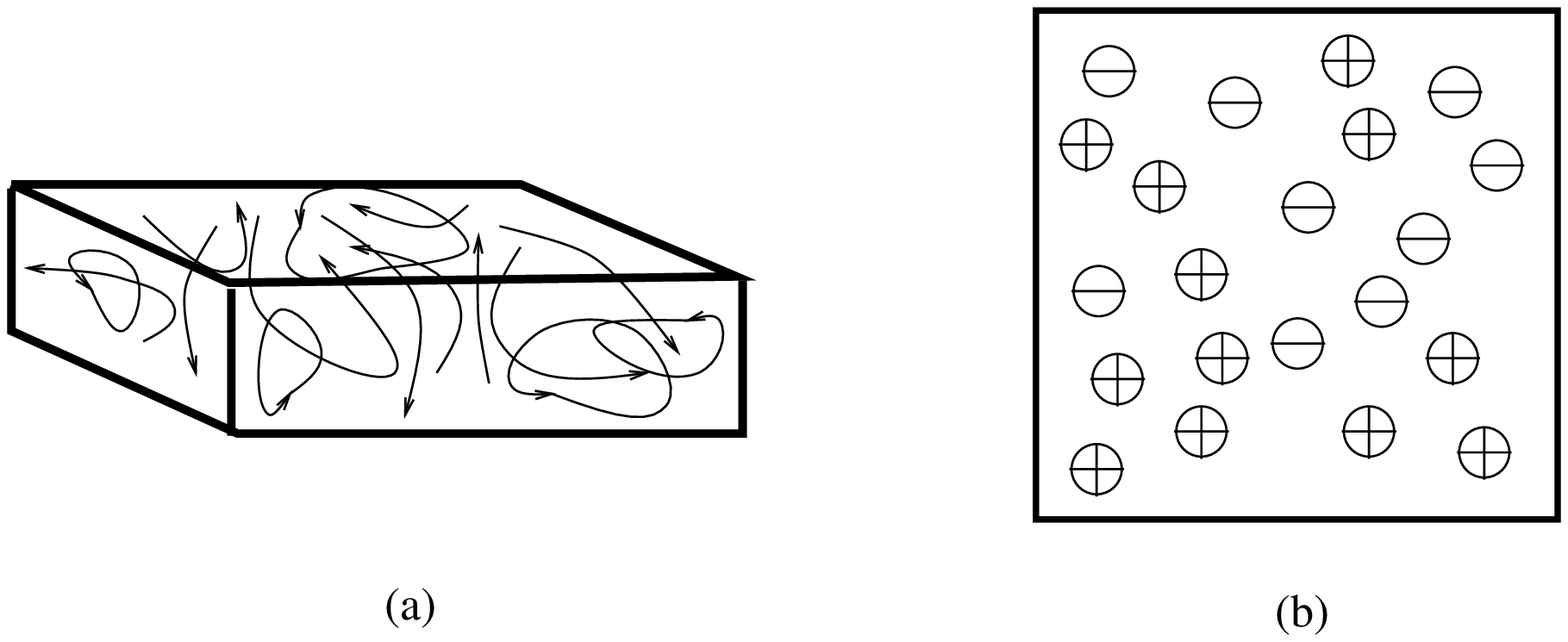}}
\vskip -2in
\end{center}
\caption{}
\label{Fig.1} 
\end{figure}
    
\newpage

\begin{figure}[h]
\begin{center}
\vskip 0.5 in
\leavevmode
\epsfysize=15truecm \vbox{\epsfbox{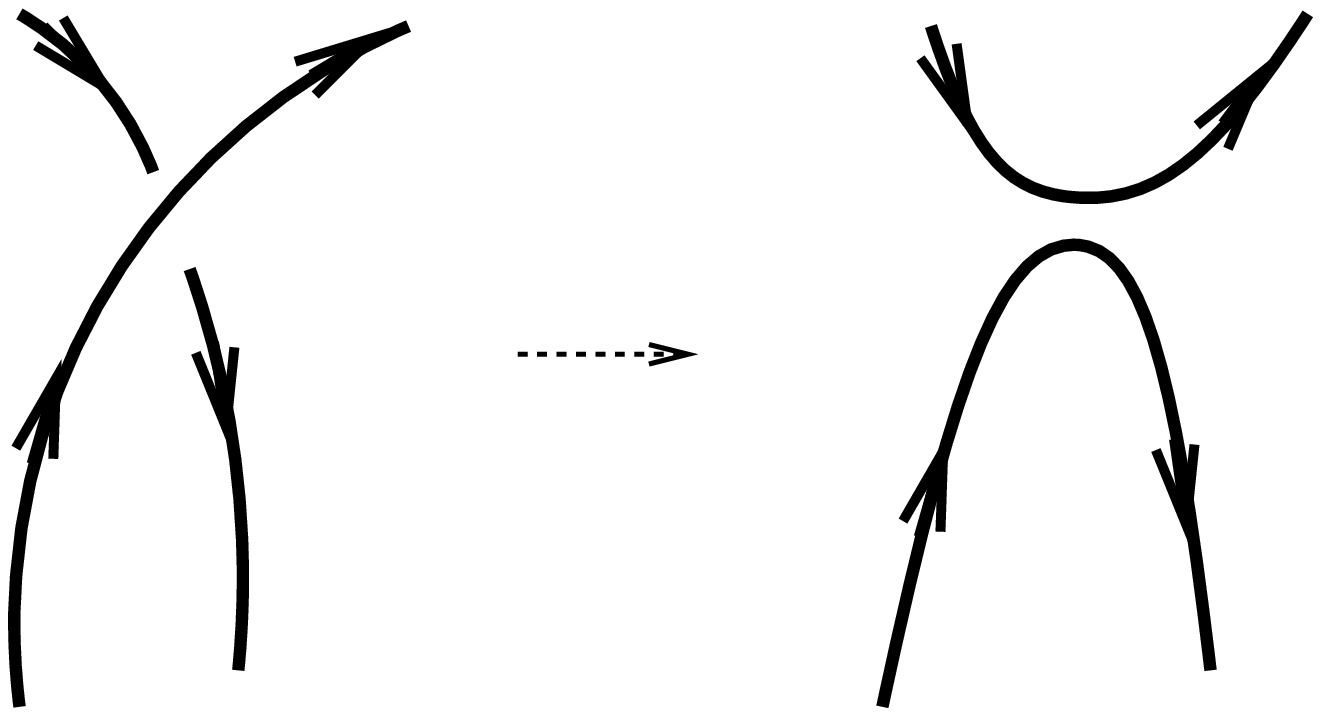}}
\vskip -1.5in
\end{center}
\caption{}
\label{Fig.2} 
\end{figure}
   
\newpage
 
\begin{figure}[h]
\begin{center}
\leavevmode
\epsfysize=15truecm \vbox{\epsfbox{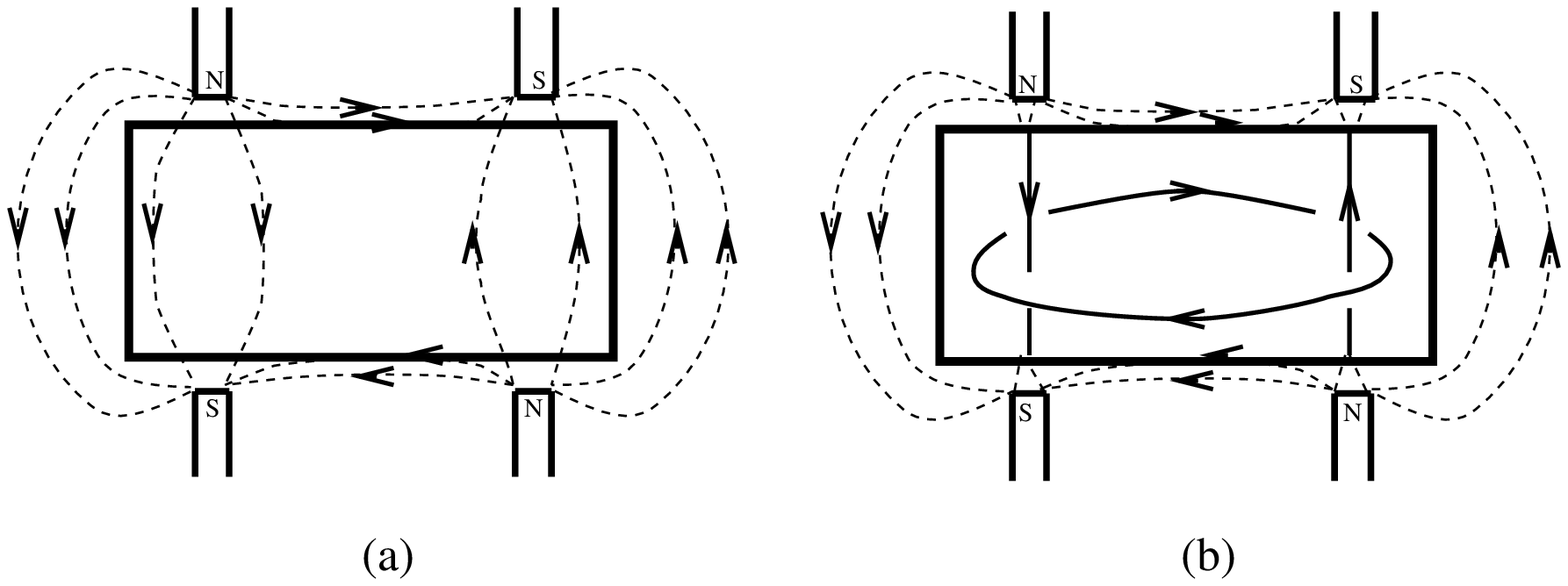}}
\vskip -1.5in
\end{center}
\end{figure}
    
\begin{figure}[h]
\begin{center}
\vskip -2 in
\leavevmode
\epsfysize=15truecm \vbox{\epsfbox{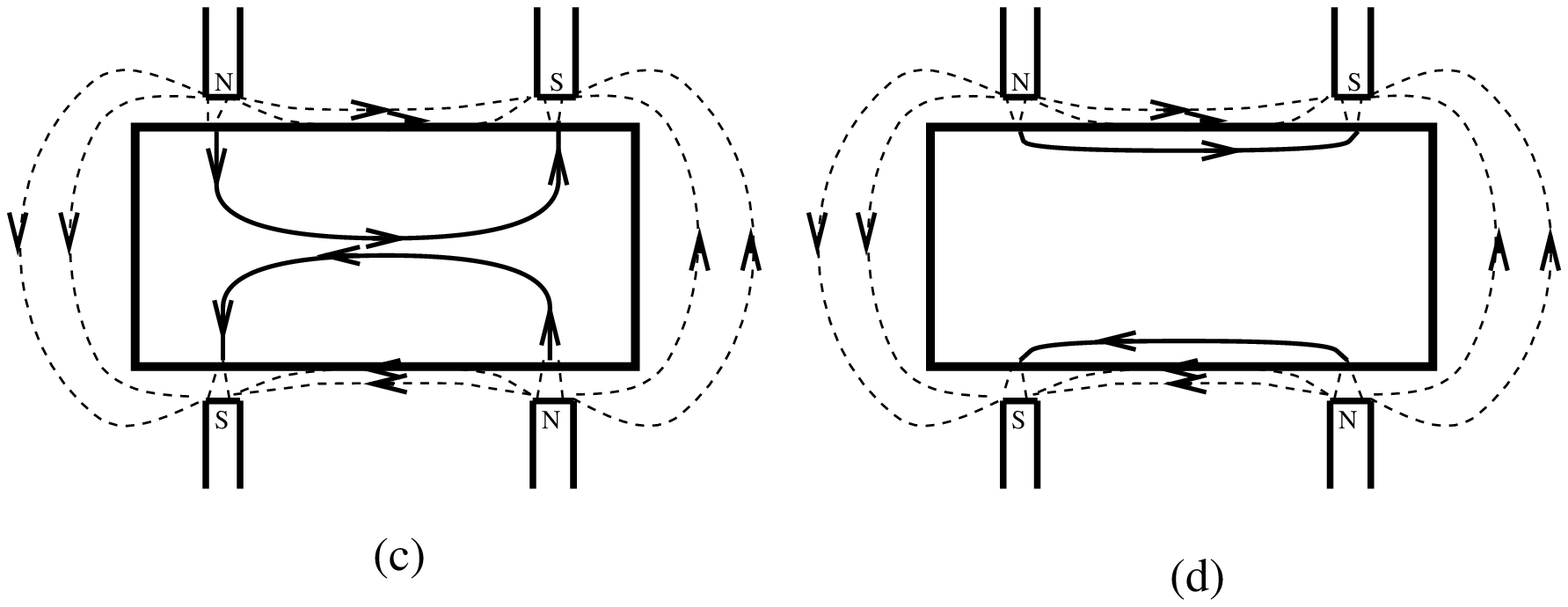}}
\vskip -2in
\end{center}
\caption{}
\label{Fig.3} 
\end{figure}
    
\newpage

\begin{figure}[h]
\begin{center}
\vskip -1 in
\leavevmode
\epsfysize=16truecm \vbox{\epsfbox{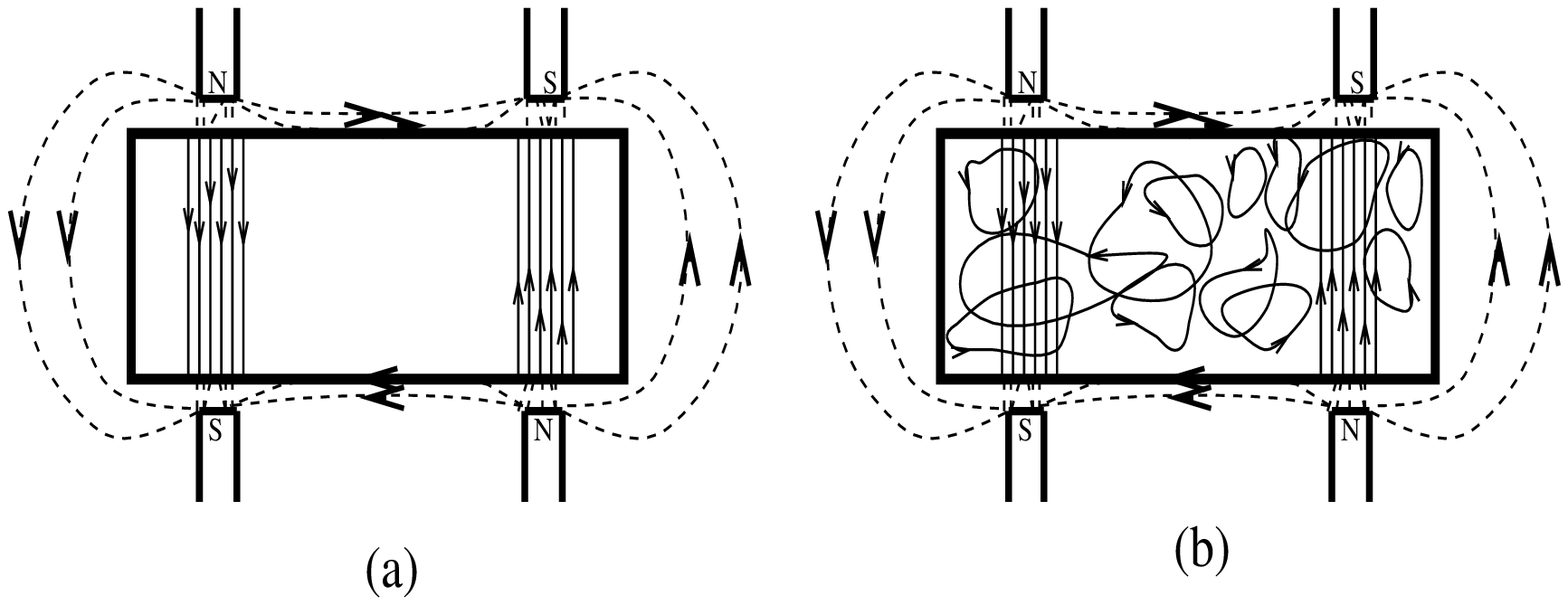}}
\vskip -1in
\end{center}
\end{figure}
    
\begin{figure}[h]
\begin{center}
\vskip -2 in
\leavevmode
\epsfysize=8truecm \vbox{\epsfbox{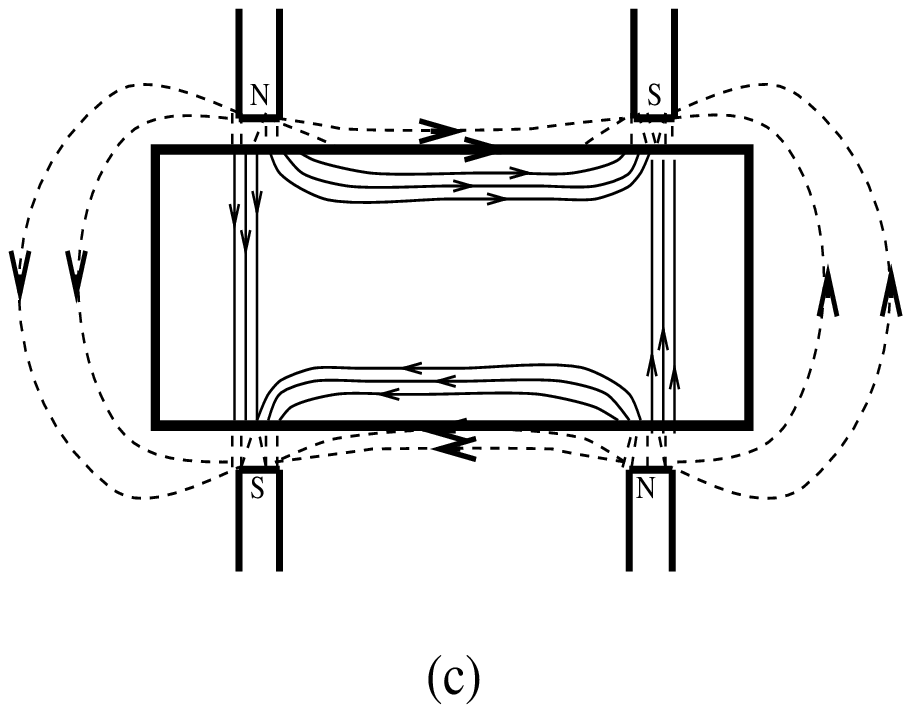}}
\vskip -1in
\end{center}
\caption{}
\label{Fig.4} 
\end{figure}


\begin{thebibliography}{99}

\bibitem{vs1} {\it Formation and interactions of topological defects},
Edited by, A.C. Davis and R. Brandenberger, Proceedings of NATO Advanced
Study Institute, 1994, (Plenum, New York).

\bibitem{vs2} A. Vilenkin and E.P.S. Shellard, {\it Cosmic strings and 
other topological defects}, (Cambridge University Press, Cambridge, 
1994); A.J. Gill, Contemp. Phys. {\bf 39}, 13 (1998).  
 
\bibitem{thrm} F. A. Bais and S. Rudaz, Nucl. Phys. {\bf B170}, 507 (1980);  
F. Liu, M. Mondello and N. Goldenfeld, Phys. Rev. Lett. {\bf 66}, 3071 
(1991).

\bibitem{kbl} T.W.B. Kibble, J. Phys. {\bf A9}, 1387 (1976).

\bibitem{dgl} S. Digal and A.M. Srivastava, Phys. Rev. Lett, {\bf 76}, 
583 (1996); S. Digal, S. Sengupta, and A.M. Srivastava, Phys.
Rev. {\bf D55}, 3824 (1997); {\it ibid} {\bf D56}, 2035 (1997).

\bibitem{cs} E.J. Copeland and P.M. Saffin, Phys. Rev. {\bf D54}, 6088 
(1996).

\bibitem{smln} T. Vachaspati and A. Vilenkin, Phys. Rev. {\bf D30},
2036 (1984).

\bibitem{nlc} M.J. Bowick, L. Chandar, E.A. Schiff and A.M. Srivastava,
Science {\bf 263}, 943 (1994).

\bibitem{zurk1} W.H. Zurek, Nature {\bf 317}, 505 (1985). See also, 
Acta Phys. Pol. {\bf B24}, 1301 (1993).

\bibitem{turok} I. Chuang, R. Durrer, N. Turok and B. Yurke, Science 
{\bf 251}, 1336 (1991).

\bibitem{he43} P.C. Hendry, N.S. Lawson, R.A.M. Lee, P.V.E. McClintock,
and C.D.H. Williams, J. Low. Temp. Phys. {\bf 93}, 1059 (1993); G.E. 
Volovik, Czech. J. Phys. {\bf 46}, 3048 (1996) Suppl. S6.

\bibitem{crln} S. Digal, R. Ray, and A.M. Srivastava, hep-ph/9805502.

\bibitem{gdsk} S. Rudaz and A.M. Srivastava, Mod. Phys. Lett. {\bf A8},
1443 (1993).

\bibitem{brnd} M. Hindmarsh, A.C. Davis and R.H. Brandenberger, Phys. 
Rev. D49 (1994) 1944; see also, R. H. Brandenberger and A.C. Davis,
Phys. Lett. {\bf B332}, 305 (1994).

\bibitem{kv} T.W.B. Kibble and A. Vilenkin, Phys. Rev. {\bf D52}, 679 
(1995); see also, J. Borrill, T.W.B. Kibble, T. Vachaspati and
A. Vilenkin, Phys. Rev. {\bf D52}, 1934 (1995). 

\bibitem{gennes} P.G. de Gennes, ``The Physics of Liquid Crystals"
(Clarendon Press, Oxford, 1974).

\bibitem{zurk2} W.H. Zurek, Phys. Rep. {\bf 276}, 177 (1996).

\bibitem{ams} A. M. Srivastava, Phys. Rev. {\bf D43}, 1047 (1991).

\bibitem{anis} S. Kolesnik, T. Skoskiewicz, J. Igalson and Z. Korczak,
Phys. Rev. {\bf B45}, 10158 (1992).

\bibitem{elec} K. Harada, T. Matsuda, J. Bonevich, M. Igarashi, S. Kondo,
G. Pozzi, U. Kawabe and A. Tonomura, Nature {\bf 360}, 51 (1992);
K. Harada, T. Matsuda, H. Kasai, J. E. Bonevich, T. Yoshida, 
U. Kawabe and A. Tonomura, Phys. Rev. Lett. {\bf 71}, 3371 (1993).

\bibitem{trp} T. J. Jackson, M. N. Keene, W. F. Vinen and P. Gilberd, 
Physica B {\bf 165\&166}, 1437 (1990).

\bibitem{shlrd} E.P.S. Shellard, Nucl. Phys. {\bf B283}, 624 (1987);
K.J.M. Moriarty, E. Myers and C. Rebbi, Phys. Lett. {\bf B207}, 411 
(1988).

\bibitem{carl} C. Rosenzweig and A.M. Srivastava, Phys. Rev. 
{\bf D43}, 4029 (1991).

\bibitem{crs} A.M. Srivastava, Phys. Lett. {\bf A194}, 141 (1994). 

\bibitem{crshtc} C. Carraro and D.S. Fisher, Phys. Rev. {\bf B51},
534 (1995); M.A. Moore and N.K. Wilkin, Phys. Rev. {\bf B50},
10294 (1994).

\bibitem{force} R. J. Adler and W. W. Anderson, J. Appl. Phys. {\bf 68},
695 (1990).

\bibitem{park} A.L. Fetter and P.C. Hohenberg in `` Superconductivity",
Vol. 2, Edited by R.D. Parks, (Marcel Dekker, Inc., New York, 1969).

\bibitem{mnfld} D.R. Nelson and H.S. Seung, Phys. Rev. {\bf B39},
9153 (1989).

\end{thebibliography}
\end{document}